\documentclass[aip,pop,twocolumn,numerical,reprint]{revtex4-2}
\usepackage{amsopn,amsmath,amssymb,amsfonts}
\usepackage{color}
\usepackage{graphicx}
\usepackage{array,tabularx}

\begin{document}


\title[Nonlinear symmetry breaking in ETG turbulence]{Nonlinear symmetry breaking in electron temperature gradient driven turbulence} 



\author{Salomon Janhunen}
\affiliation{Oden Institute for Computational Engineering and Sciences, University of Texas at Austin, Austin, TX 78712, USA}
%
\author{Gabriele Merlo}
\affiliation{Oden Institute for Computational Engineering and Sciences, University of Texas at Austin, Austin, TX 78712, USA}
%
\author{Frank Jenko}
\affiliation{Oden Institute for Computational Engineering and Sciences, University of Texas at Austin, Austin, TX 78712, USA}
\affiliation{Max Planck Institute for Plasma Physics, Garching, Germany}
\author{Alexey Gurchenko}
\affiliation{Ioffe Institute, 194021 St. Petersburg, Russia}
\author{Evgeniy Gusakov}
\affiliation{Ioffe Institute, 194021 St. Petersburg, Russia}
\author{Timo Kiviniemi}
\affiliation{Aalto University School of Science, Espoo, Finland}


\date{\today}

\begin{abstract}
Nonlinear symmetry breaking may occur in systems with two or more states whose linear dynamics displays certain symmetries, one of which is preferred nonlinearly. We have identified a regime of electron temperature gradient (ETG) instabilities in a tokamak plasma with circular concentric flux surfaces that has its largest growth rate at a finite ballooning angle, establishing a symmetry that is nonlinearly broken to favor one sign for the ballooning angle. This is the first example of nonlinear symmetry breaking in simulations of a drift instability in the absence of externally imposed flow shear or asymmetry in the plasma column.
\end{abstract}


\maketitle 

\section{Introduction}


The electron temperature gradient mode can act as a driver of experimentally relevant levels of electron heat transport, despite its small characteristic spatio-temporal scales. A rich field of nonlinear physics has been identified within electron temperature gradient (ETG) mode turbulence, driven by the ETG instability\cite{Lee1987,Horton1988,Hirose1990,Dong2002,Smolyakov2002}. ETG turbulence is characterized by long streamers that enhance transport to levels exceeding quasi-linear estimates,\cite{JenkoPoP2001,Dorland2000,Jenko2001,JenkoPRL2002,Gurcan2004} and simplified models of ETG turbulence admit soliton-like solutions\cite{Gurcan2004,Khan2017}. Self-driven zonal flows are in most situations too weak to saturate ETG turbulence, so saturation happens through secondary instabilities of the streamer structures.\cite{JenkoPoP2001,Dorland2000,JenkoPRL2002,LiPoP2005} Inverse cascades provide an effective mechanism for distributing ETG energy to larger scales, with the effect that ETG turbulence may affect turbulence and zonal flows arising from ion scale instabilities.\cite{Howard2015} Cross-scale interactions between ion scale instabilities and ETG as well as zonal flow dynamics can complicate the picture of ETG saturation as well.\cite{Smolyakov2002,JenkoJPFR2004,GoerlerPRL2008,Asahi2014,MaeyamaPRL2015,HowardPoP2016}

Nonlinear symmetry breaking is a phenomenon observed in many kinds of systems, ranging from quantum to classical. Here, the ETG instability exhibits growth at a finite ballooning angle for parameters used in this paper, giving rise to symmetry with respect to the radial wave number $k_r$. In nonlinear simulations, however, the amplitude spectrum of density fluctuations concentrates on one sign of ballooning angle even though both signs have equal growth rates in linear simulations. As we will proceed to show below, this process is robust and depends on the relative directions of the toroidal magnetic field and plasma current.

Maximal growth rates at finite ballooning angle have been seen before in other simulations of drift-wave turbulence\cite{JenkoPoP2009,Camenen2011,Migliano2013,SinghPOP2014,Xie2016,LuPoP2017,KaangPoP2018,LuPPCF2019,parisi2020toroidal}.
Generally in the nonlinear state for gyrokinetic turbulence simulations the presence of inclined fluctuations have not been seen to be of major significance before, perhaps because they tend to occur at higher $k$ and therefore contribute less to transport. Also, symmetry in many examples is commonly broken already for the linear eigenmodes through up-down asymmetry of the geometry or a priori imposed flows, but significant asymmetry in fully developed turbulence is not seen.\cite{JenkoPoP2009} Our system is interesting because not only is symmetry broken in an up-down symmetric geometry in the absence of externally imposed flows, broken symmetry is retained in the turbulent spectrum in long term simulations. This suggests that the reason for broken symmetry lies within something more fundamental than simply a choice of configuration.

The structure of the paper proceeds as follows: we introduce the physical parameters derived from experiments at the FT-2 tokamak used for simulations performed with GENE and discuss features of GENE in simulating the high-$k$ turbulence that arises from this set-up. We show how for the parameters at the upper-hybrid resonance layer for enhanced scattering measurements maximal growth rates of electron temperature gradient modes occur at a non-zero ballooning angle establishing a symmetry where breaking may occur. We illustrate how symmetry is broken in nonlinear simulations and sustained in long-term simulation of ETG turbulence; establish that symmetry is broken in a manner that depends on the relative directions of toroidal magnetic field and current, and finally discuss an analogous problem of ion temperature gradient turbulence where such symmetry breaking does not appear and give an example of trapped electron mode turbulence where broken symmetry is also apparently absent. We conclude the report with discussion of the salient features in our simulations as well as a summary of the results.

\section{Methods and techniques}

GENE (\url{http://www.genecode.org} serves as an extensive reference) is a gyrokinetic Eulerian Vlasov code that solves the electromagnetic gyrokinetic equations in field-following coordinates. It has been used extensively in various investigations into tokamak plasma turbulence and beyond.\cite{JenkoPoP2001,JenkoPRL2002,DannertPoP2005,GoerlerPRL2008,PueschelAPJ2014,ToldPRL2015} For the present work, notable features of GENE are the possibility to run global or flux tube simulations, a robust collision model that allows investigations to reasonably high collisionalities present at the FT-2 tokamak\cite{Gusakov_2006a}, and several gyrokinetic species that interact through self-consistently solved electromagnetic fields.

GENE uses a field line following coordinate system $(x,y,z)$ for discretization of the distribution function $f(x,y,z,v_\parallel,\mu)$ and the fields $\phi(x,y,z)$ and $A_\parallel(x,y,z)$. The radial coordinate $x$ acts as the flux-surface label, the binormal coordinate $y$ is tangential to the flux-surface and normal to the magnetic field line, and $z$ is a poloidal angle used as the field line label in the parallel direction. The distribution function and fields are defined using a discrete Fourier basis in $y$, leading to a discretization in the binormal wave number $k_y$. 

There are myriad ways to perform simulations using GENE, such as local flux tube in both circular and Miller geometry, to fully global including an arbitrary number of particle species within a magnetic equilibrium numerically defined by solving the Grad-Shafranov equation. In this paper we use GENE primarily in local flux tube mode using a circular equilibrium, where a radially periodic box is defined with a discrete Fourier basis in $x$, also giving the wave number $k_x$.

The case at hand is based on experiments performed on FT-2 tokamak at Ioffe Institute, Russia.\cite{Gusakov_2006b,Gurchenko_2007}. In the preceding paper, the electron density profile has been obtained from Thomson scattering and interferometry, and the electron temperature profile is obtained from Thomson scattering. The main ion temperature profile was measured by neutral particle analyzer and spectroscopy. The equilibrium was reconstructed from experimental data using ASTRA modeling. There are three main species in the experiment: electrons, protons and fully ionized oxygen O$^{8+}$ impurity. Electron density fluctuations are measured using {\it enhanced scattering} from the upper hybrid resonance layer\cite{Gusakov_2006a,Gusakov_2006b,Gurchenko_2007}, showing a distinct component at wave lengths corresponding to ETG turbulence\cite{Gurchenko2010}.

Because of the short wave length nature ETG turbulence, we proceed with numerical analysis of the experimental situation using local flux tube GENE simulations centered at the upper hybrid resonance layer where the experimental measurements are performed in Refs.~\onlinecite{Gusakov_2006a,Gusakov_2006b,Gurchenko_2007,Gurchenko2010}. Parameters for the simulations are given in Table \ref{tab:params}. Gradients are defined via the parameters $\omega_{n,T}=R_0/L_{n,t}$ where $L_n=|\nabla\log n|^{-1}$ and $L_T=|\nabla\log T|^{-1}$.

GENE is used as an initial value solver for determining the linear growth rates and frequencies for toroidal modes, and nonlinear simulations are performed for characterizing the turbulence that arises from the growth of the linear modes. Linear growth rates are calculated in GENE for a single $k_y$ mode at a time giving the fastest growing mode for that particular value of $k_y$, and scanning over $k_y$ to obtain the growth rate and frequency spectra. Each simulation for $k_y$ includes a narrow band of $k_x$ modes, so to calculate the modes that have a finite ballooning angle $\vartheta_0$ the $k_x$ offset $k_{x0}$ needs to be scanned over as well. 

Numerical parameters used for the ETG case in nonlinear simulations as presented in this paper are as follows: the simulation box is $(L_x,L_y,L_{v_{\shortparallel}},L_{\mu})=(10.8,25,3,9)$, resolution is $(N_x,N_y,N_z,N_{v_{\shortparallel}},N_{\mu})=(256,96,24,45,15)$. The simulation was repeated with different boxes and resolutions: up to $L_x=40$ with $N_x=1024$; up to $N_y=128$; using $N_z=64$; and some minor variations on $N_{\mu}$ and $N_{v_{\shortparallel}}$ using different $L_{v_{\shortparallel},\mu}$ boxes. Lengths are normalized to $\rho_s$.

\begin{table}
  \caption{Local parameters for GENE simulations of ETG turbulence.}\label{tab:params}
  \begin{tabularx}{0.8\columnwidth} { 
  | >{\raggedright\arraybackslash}X 
  | >{\centering\arraybackslash}X | }
    \hline
    R$_0$ [m] & 0.55\\
    a  [m]& 0.08\\
    B$_0$ [T] & 2.2\\
    q$_0$ & 2.4\\
    $\hat{s}$ & 1.1\\
    $\omega_{ne}$  & 15.45 \\
    $\omega_{Te}$  & 21.72 \\
    $\omega_{T_H,T_{O^{8+}}}$ & 15.6 \\
    $T_e$ [keV] & 0.1664\\
    $n_e$ [10$^{19}$/m$^3$]& 1.581\\
    Z$_{\text{eff}}$ & 3\\
    $\tau$ & 7.108\\
    $r/a$ & 0.7\\
    \hline
  \end{tabularx}
\end{table}

Spectra for turbulence are show in units of $k_{x,y}\rho_s$, with $\rho_s=c_s/\Omega_i$ where $c_s^2=T_e/m_i$ is the ion-acoustic velocity, $T_e$ is electron temperature, $\Omega_i=q_e B/m_i$, $q_e$ being the unit charge, $B_0$ is the on-axis magnetic field, $m_i=1\,$amu the hydrogen mass. Real electron-ion mass ratio is used where required, and the impurity species is taken to be $\text{O}^{8+}$.

Fluctuation spectra are calculated from the simulation using time averaging to obtain a RMS $S(n_e)^2=\int_{t0}^{t1} |n_e(k_x,k_y,z;t)|^2 \,dt/(t_1-t_0)$, evaluated at $z=0$.

We will concentrate on the linear analysis in the next section.

\section{Linear growth rate spectrum symmetric in $\vartheta_0$}

A simplified version of the equations solved by GENE can be cast into a linear dispersion relation for ETG as\cite{Hirose1990}
\begin{equation}
  \epsilon \left( \omega ,\mathbf{k}\right) =-\frac{k_\perp^2}{\lambda_{De}^2}+\sum_a \epsilon_{a}\left( \omega ,\mathbf{k}\right)=0,
\label{disp}
\end{equation}
where $\epsilon_{a}$ are the susceptibilities from electrons ($a\rightarrow e$), ions ($a \rightarrow i$) and impurities ($a \rightarrow Z$)
\begin{equation}
  \epsilon_{a}\left(\omega,\mathbf{k}\right)=\int\frac{q_a\phi}{T_a}f_{Me}+\frac{\omega+\omega_*}{\omega-\omega_{Da}-k_{\parallel} v_\parallel}J_0^2(b_a)\frac{q_a \phi}{T_a} f_{Ma}\,d^3v,
\end{equation}
where $b_a=k_{\perp}^{2}\rho _{a}^{2}$, $\rho _{a}^{2}=v_{a}^{2}/\Omega _{ca}^{2}$ is the Larmor radius of species $a$,
$v_{a}^{2}=T_{a}/m_{a}$, $\lambda_{De}^{2}=\frac{\epsilon_{0}T_{e}}{n_{0}q_{e}^{2}}$ is the Debye length, $J_{0}(x)$ is the Bessel function of the 1st kind, $k_\perp\equiv\left\vert \mathbf{k}\right\vert =\sqrt{k_{x}^{2}+k_{y}^{2}}$ is the perpendicular wave number. ETG waves are slow compared to the electron gyration period, so the Bessel series has been truncated at the first term (in contrast to Ref.~\onlinecite{JanhunenPOP2018}). For short wavelength ETG modes we can approximate the ion responses to be adiabatic (with $J_0^2(b_a)\rightarrow 0$ as $b_a$ becomes large)\cite{Reshko_2008}, giving the dispersion relation
\begin{equation}\label{eq:fulldispersion}
\left(1+\tau\right)\lambda_{De}^2+k^2=\int\frac{\omega+\omega_{*e}}{\omega-\omega_{De}-k_{\parallel} v_\parallel}J_0^2(b_e) f_{Ma}\,d^3v,
\end{equation}
where $\tau=Z_{\text{eff}}T_e/T_i$ is a factor for the adiabatic response of the ions, $Z_{\text{eff}}=(Z_{Z}^2n_z+n_i)/n_e$ is the effective ion charge obtained from quasineutrality, with $Z_Z$ denoting the charge state of the impurity species. Drift frequencies $\omega_D=\vec{k}\cdot\vec{v}_D$ and $\omega_*=\vec{k}\cdot\vec{v}_*$ correspond to the magnetic and diamagnetic drifts
\begin{equation}\label{eq:drifts}
  \begin{gathered}
   \vec{v}_{Da}=b\times\left(\frac{v_\parallel^2}{\Omega_a} b\cdot\nabla b+\frac{v_\perp^2}{2 \Omega_a} \nabla\ln B\right),\\
   \vec{v}_{*a}=\frac{b\times\nabla p_a}{m_a n_a\Omega_a}.
  \end{gathered}
\end{equation}
In Eq.~(\ref{eq:fulldispersion}) the right hand side corresponds to the non-adiabatic electron response, where the diamagnetic drift acts in the binormal direction (in our notation, coordinate $y$) because the equilibrium pressure is a flux function to lowest order, but the magnetic drift has also a radial component (coordinate direction $x$) which contributes in the dispersion relation. We may write $k_x=\hat{s}k_y\left(y-\vartheta_0\right)$, where $\vartheta_0=k_{x0}/\hat{s}k_y$ is the ballooning angle. It is important to note that the value of $\vartheta_0$ will be dependent on $k_y$ in the linear simulations, so its introduction has not reduced generality of our model at all. Frequently largest growth occurs at $\vartheta_0=0$ corresponding to maximum amplitude at the outboard of the torus, which motivates its wide use in analysis of drift waves.

\begin{figure}[htp]
\begin{center}
  \includegraphics[height=2in,clip]{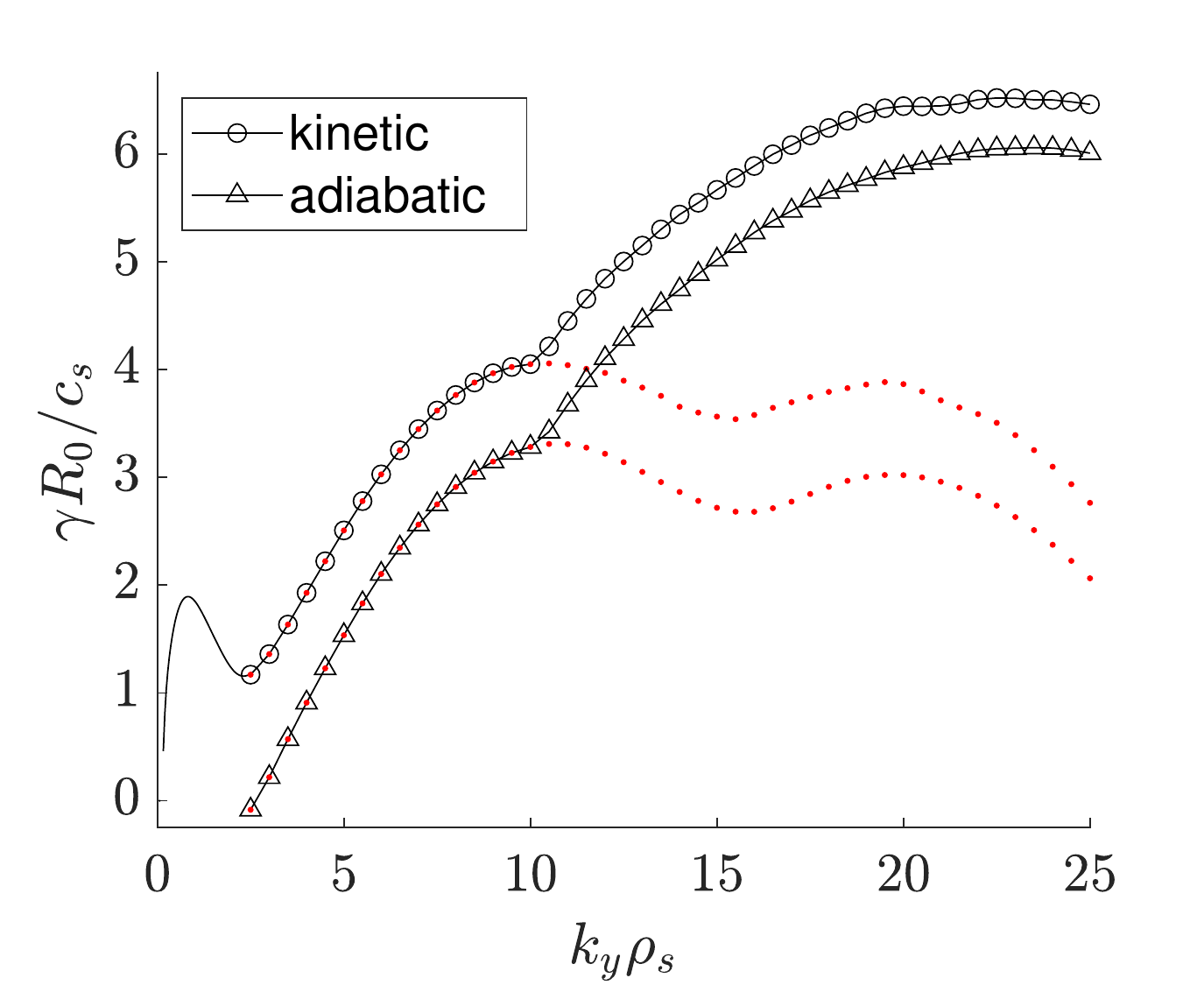}%
  \hspace{0.25in}
  \includegraphics[height=2in,clip]{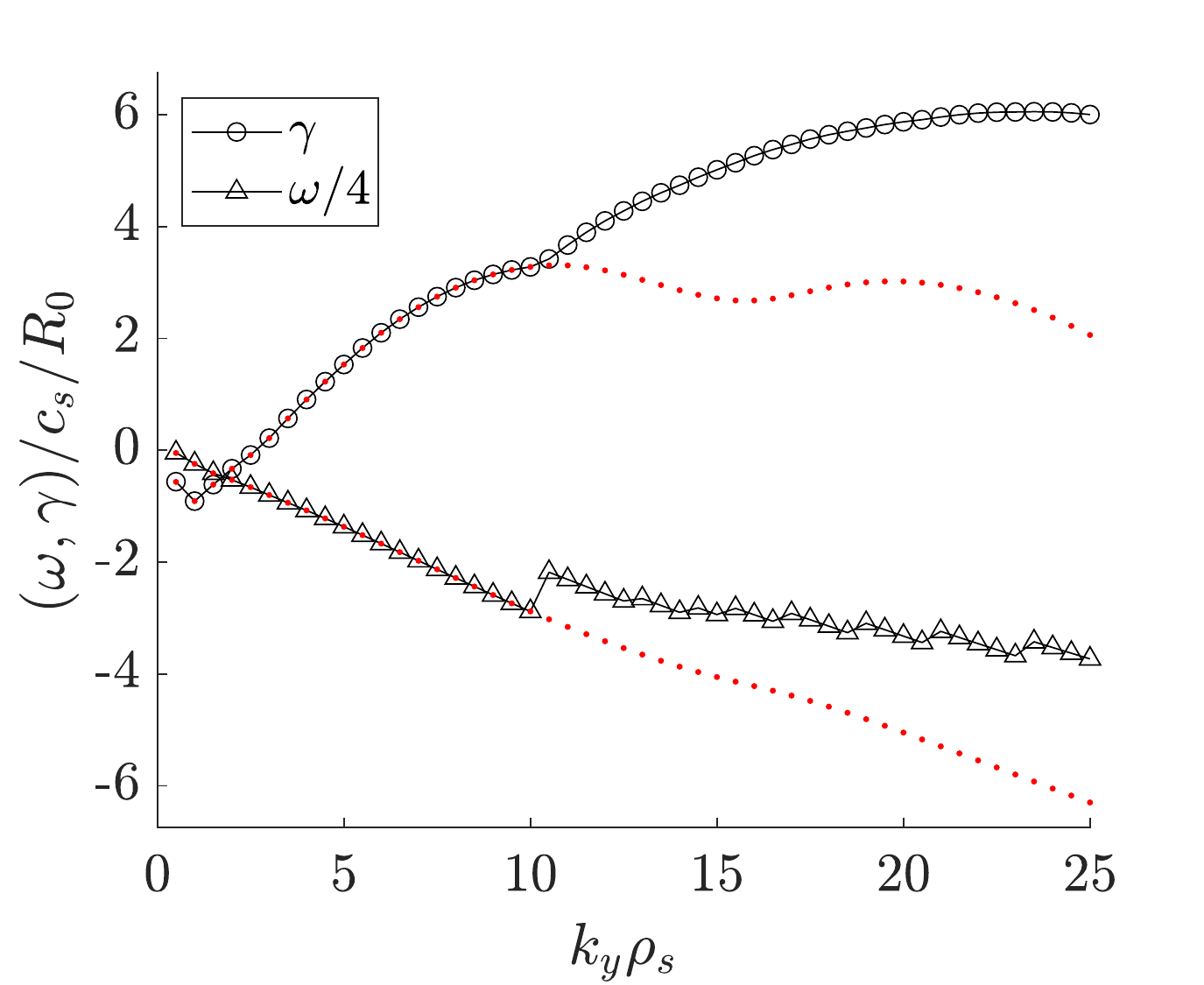}%
\end{center}
\caption{Left: growth rate across ion and electron scales from simulations with kinetic ions ($\bigcirc$) and adiabatic ions ($\triangle$). Kinetic simulation is with hydrogen and $\text{O}^{8+}$ ions giving the same $\tau$ and $Z_{\text{eff}}$, both performed for the flux-tube centered at $r/a=0.7$ where the measurements have been made. Dotted red line shows the growth rate at $k_x=0$. Right: growth rate ($\bigcirc$) and frequency ($\triangle$) at $\text{max}\,\gamma(k_x)$ for the adiabatic case. Dotted red lines show growth rates and frequencies at $k_x=0$.\label{fig:etg-omegamma}}
\end{figure}

\begin{figure}[htp]
\includegraphics[width=\columnwidth,clip]{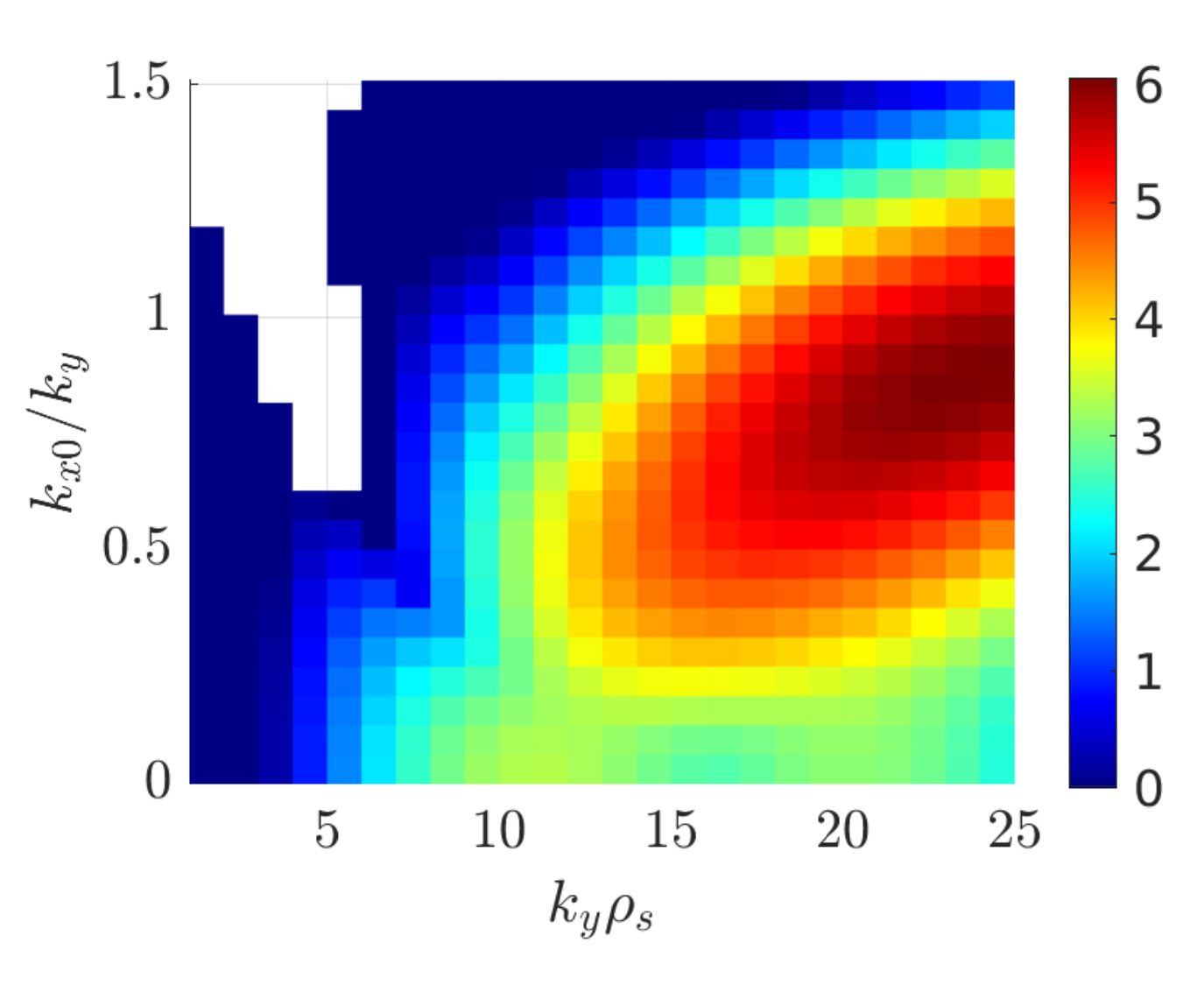}%
\caption{Growth rate as the ballooning angle is varied in the simulation with adiabatic ions. Here we represent the ratio of the $k_x$ offset $k_{x_{0}}$ to the $k_y$ as a proxy for the ballooning angle. \label{fig:etg-omegamma-surf}}%
\end{figure}

A finite ballooning angle becomes important for ETG modes at higher $k_y$ in the case defined by Table \ref{tab:params} as illustrated in Fig.~\ref{fig:etg-omegamma}. The spectrum is symmetric with respect to the sign of $k_{x0}$ up to output precision, so the negative branch is suppressed in Figs. \ref{fig:etg-omegamma} and \ref{fig:etg-omegamma-surf}.

In Fig. \ref{fig:etg-omegamma} we show growth rate for the experimentally motivated case in Table~\ref{tab:params}, background density gradients for all species being equal due to quasineutrality. In the case with adiabatic ions the ion response is replaced by $\tau$ and $Z_{eff}$ in the field equation and collisions, respectively. 
There is a slight shift in the growth rates that are caused by the ion temperature gradient contribution in the kinetic case; increasing $\omega_{T}$ for the ions to equal the electron one brings the growth rates between kinetic and adiabatic cases closer whereas taking the ion temperature gradient towards zero increases the difference at the ETG scale. The trapped electron mode root vanishes for adiabatic ions, and only the ETG instability remains.

The $\vartheta_0$ symmetry manifest for the ETG modes is broken in the nonlinear regime, as we show in the following section.

\section{Symmetry breaking in nonlinear simulations}

We investigate nonlinear simulation of the case defined in Table~\ref{tab:params} using adiabatic ions due to computational expense. The simulation is initialized with a density fluctuation spectrum proportional to $k^{-1}$ at amplitude $10^{-4}$ relative to the background value. The modes are given random phases to establish fluctuations over all of the simulation region in real space. The question whether symmetry breaking is spontaenous was investigated by initializing the simulation with different initial conditions.

\begin{figure}[htp]
\includegraphics[width=0.75\columnwidth,viewport=0 0 350 366,clip]{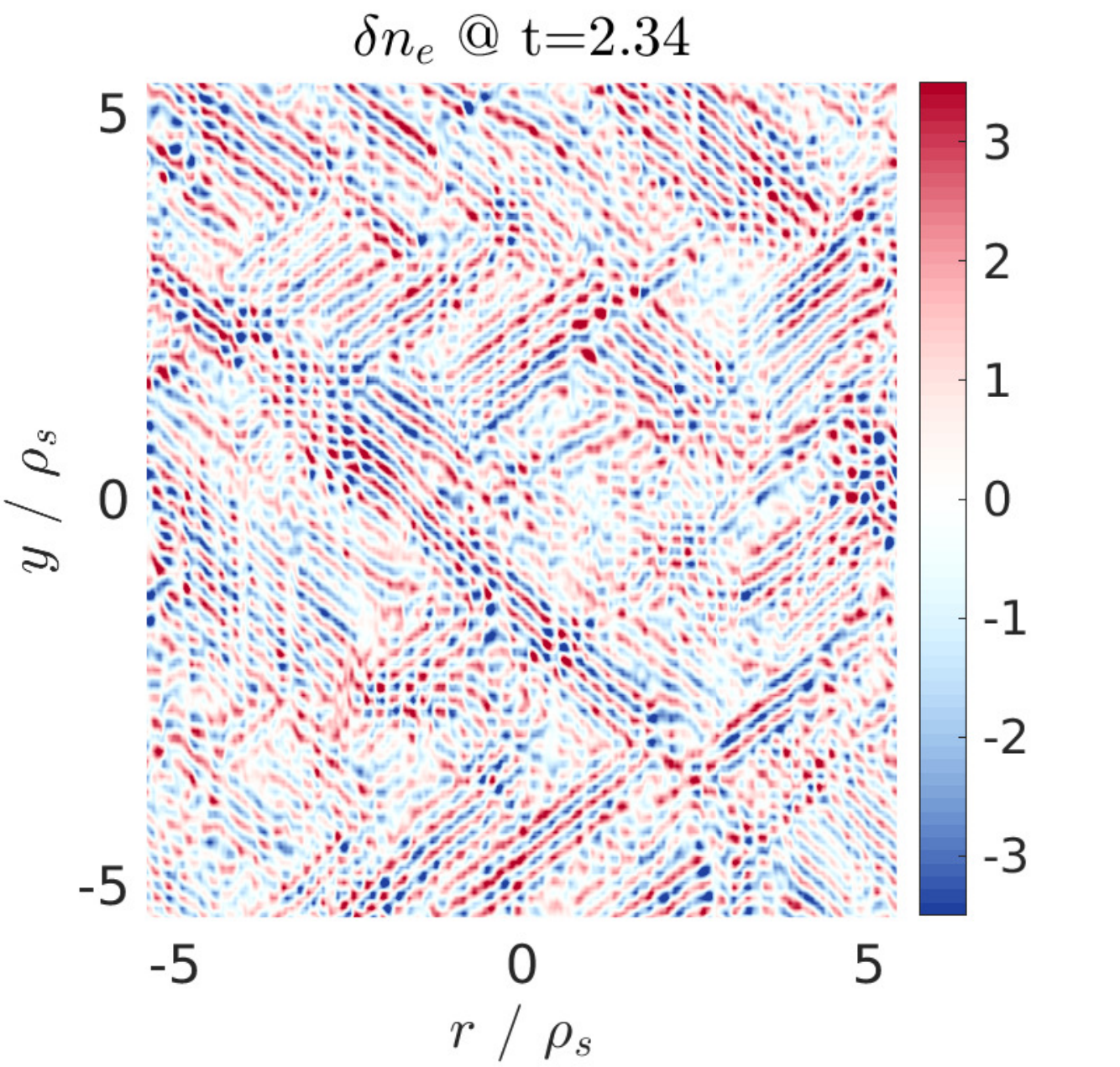}%
\hspace{5pt}
\includegraphics[width=0.75\columnwidth,viewport=0 0 350 366,clip]{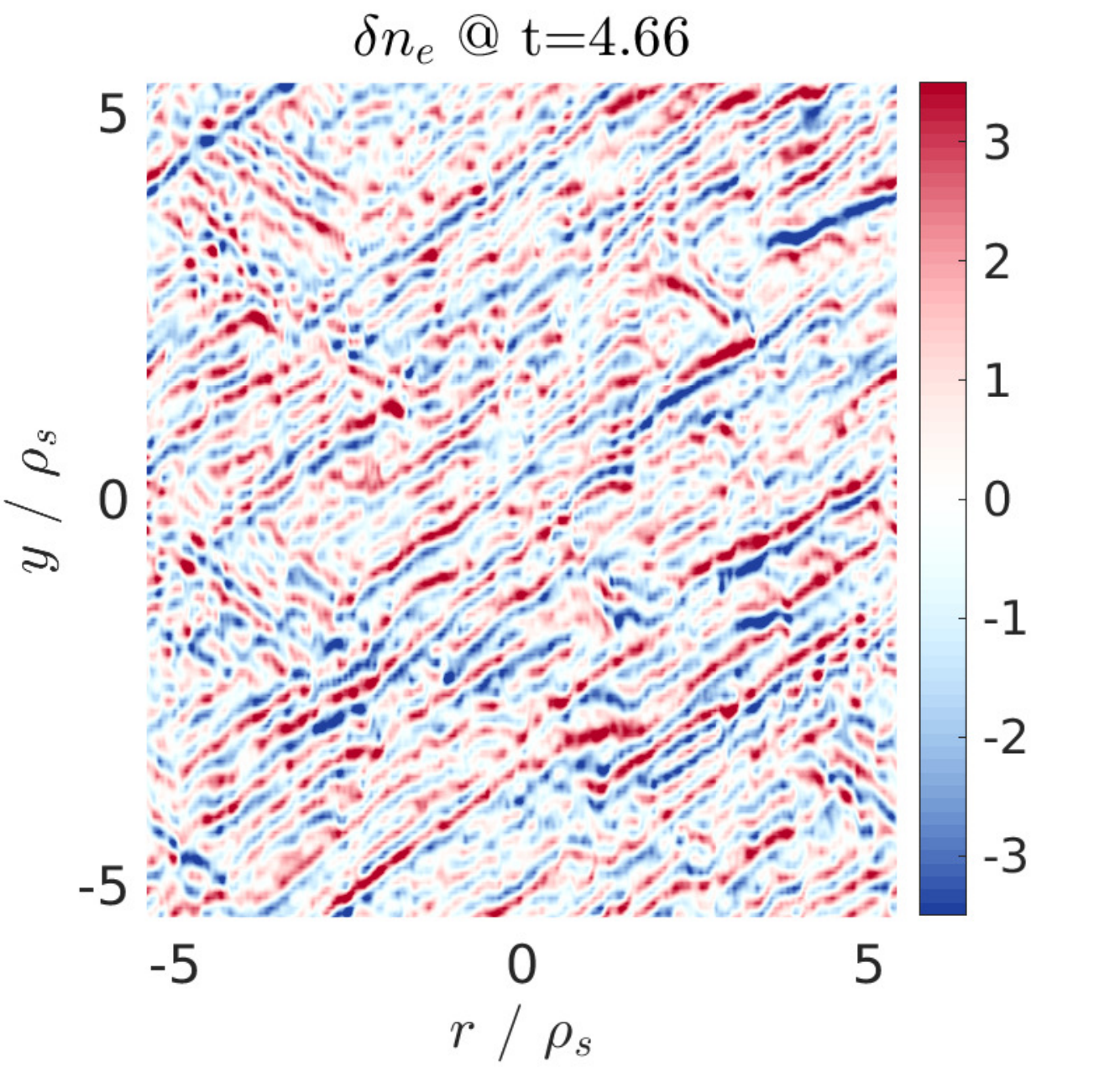}%
\hspace{5pt}
\includegraphics[width=0.75\columnwidth,viewport=0 0 350 366, clip]{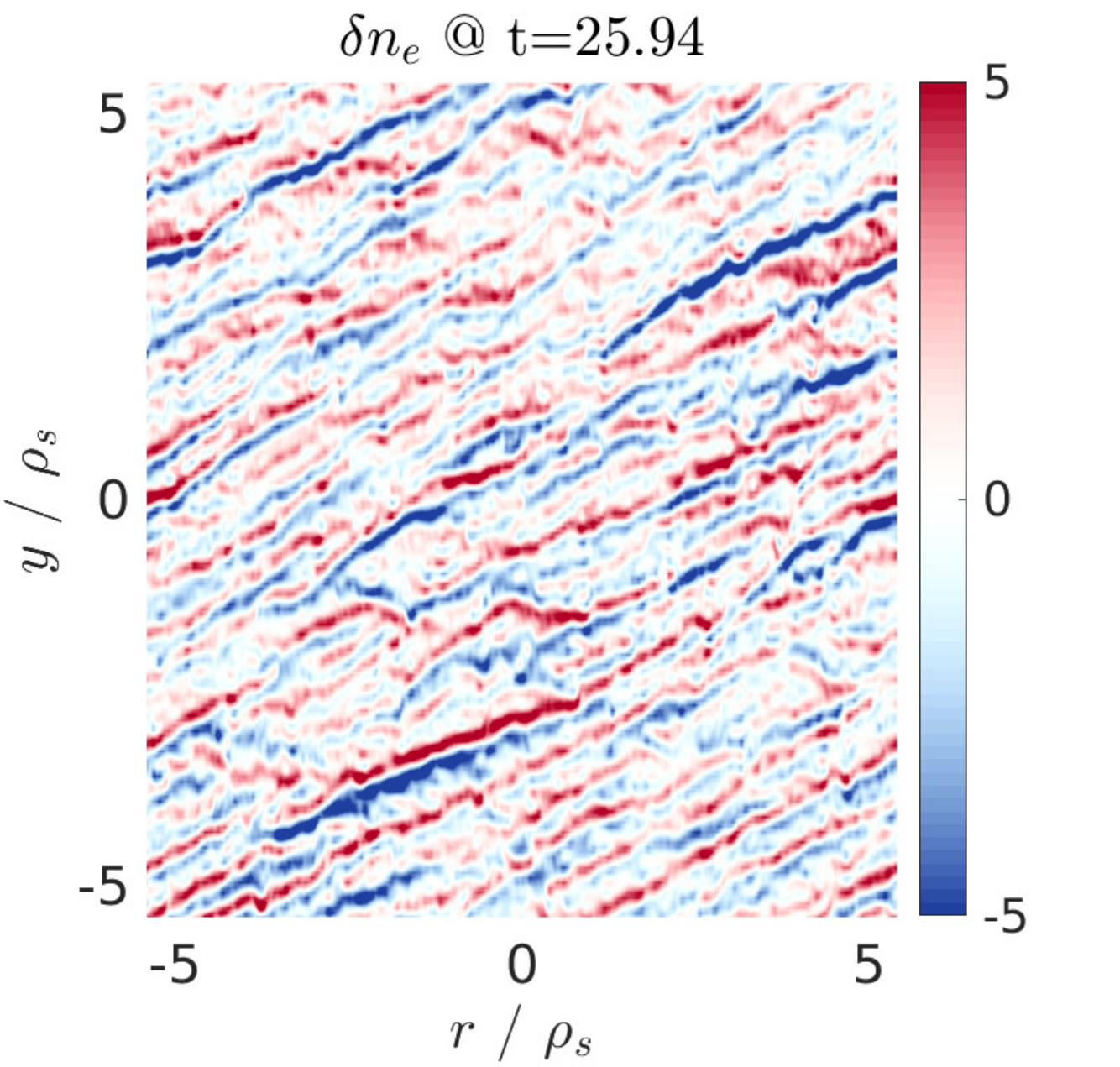}%
\caption{Contours of electron density fluctuations on a plane in the case with adiabatic ions at the early stages of the simulation.  Time is shown in units of $R_0/c_s$. }\label{fig:etg-density-contours}%
\end{figure}

The evolution of the simulation is illustrated in Fig.~\ref{fig:etg-density-contours}. Initially the fastest growing modes (with both positive and negative $k_x$) get established, and a transient peak of the heat flux occurs at $t=2.34 R_0/c_s$. Soon after ($t=4.66 R_0/c_s$) modes with negative sign gain in amplitude while positive modes diminish, and the spectrum shifts down in $k_y$. In the longer term (representative snapshot at $t=25.94 R_0/c_s$) the heat flux remains steady while slightly fluctuating about a mean value, and long streamer-like structures typically seen in ETG turbulence are observed at negative $k_x$. As the energy content progresses towards lower $k_y$ values due to inverse cascade, the mean ballooning angle of the modes decreases as well, while remaining finite. This is seen as a reduction in inclination for the fluctuations in Fig.~\ref{fig:etg-density-contours}. 


Transport saturates much earlier than the nonlinearly established fluctuation spectrum in density. It should be noted that for the present FT-2 case the vast majority (>99\%) of the heat flux is transported in trapped electron mode turbulence when kinetic ions are included and simulations are performed at the ion scales. The electron flux at the ion scales is computed as $Q_e=69\,\text{kW/m}^2$, which is of the order of the ASTRA modeling flux given by $Q_e=57\,\text{kW/m}^2$.

\subsection{Long-term nonlinear evolution}

\begin{figure}[htp]
    \centering
    \includegraphics[width=\columnwidth,clip]{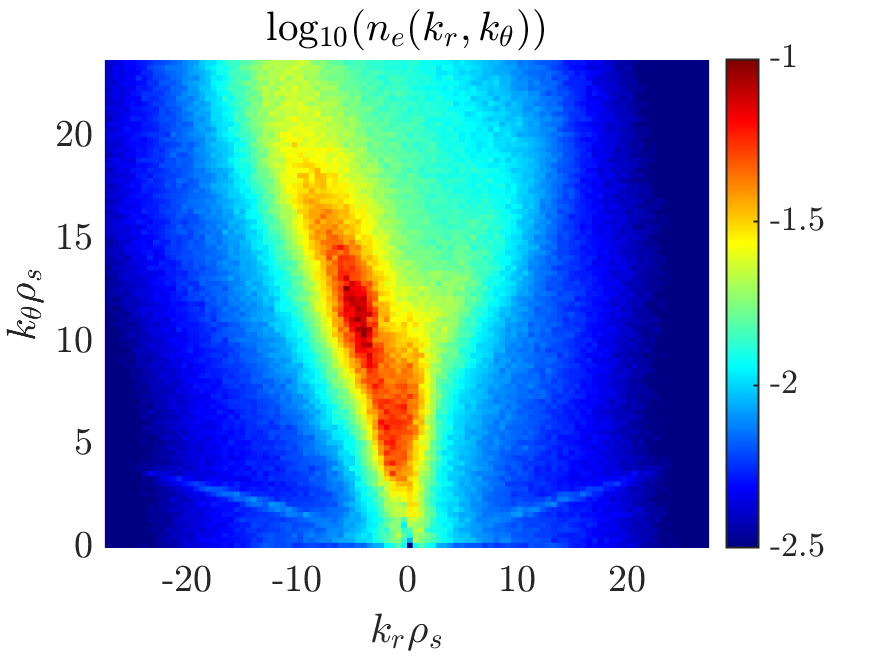}
    \caption{Spectrum of electron density fluctuations in a long-term simulation. The range of $k_x$ is reduced from the computation (which has $|k_x|\leq 60$) for better illustration.}
    \label{fig:electron_spectrum}
\end{figure}

The state of broken symmetry is a robust feature of this case, and is
sustained in the long term simulations over 100's of $R_0/c_s$. The time-averaged spectrum of density fluctuations over $50R_0/c_s$ after $200R_0/c_s$ in Fig. \ref{fig:electron_spectrum} shows the anisotropic nature of broken-symmetry streamers in ETG turbulence for the present case. The mean preferred ballooning angle can be changed by reversing to counter current in the analytical equilibrium, namely by changing $B_0 \rightarrow -B_0$ or $j_T \rightarrow -j_T$. The nonlinear spectrum is symmetrically mirrored over $k_x$ when product of the signs is negative, but linearly the growth rates and frequencies are not affected. Self-induced shearing is ineffective in suppressing the spectral asymmetry; in fact shearing appears to be highly non-correlated in time and occurs over all scales in the spectrum. The dynamics of the spectrum is unaffected by suppression of the zonal modes, showing that they are unimportant.

A small additional externally imposed $\omega_{E\times B}=10^{-4}$ toroidal shearing rate is required to stabilize a spurious low-$k_y$ mode that short-circuits the transport after $100R_0/c_s$ in the current case or a longer period if radial extent of the simulation box is increased. This happens likely due to inverse cascade, and in the absence of shearing the system prefers a low-$k_y$ streamer. The externally imposed shear rate is much lower than the growth rates in the unstable spectrum, and is of the order of neoclassical shear for these profiles. Sign of the $E\times B$ shear does not affect the broken symmetry, and introduction of the shear allows sustainment of the broken symmetry state virtually indefinitely in the simulation. Hyper-viscosity at rate $\alpha_{x,y}=0.005$ with the profile $k^4$ is used avoid pileup of the fluctuations at the resolution boundary, as well as $\alpha_z=5$ in the parallel direction. The results were very robust with respect to variations in numerical parameters, as long as the box size and resolution were sufficient.

We investigated robustness of the asymmetrical turbulent spectrum also by reflecting the $x$-coordinate for the GENE restart in the deeply nonlinear stage. Even with this highly forced initial condition the simulation recovered the original spectrum for that choice of $B_0$ and $j_T$ after a period of dynamic restructuring.

\section{Is symmetry broken in other kinds of turbulence?}

Analogies between different drift wave turbulence types can be made. Simplified turbulence models such as the Hasegawa-Mima-Charney equation are analogous for ETG and ITG turbulence, except how zonal modes affect turbulence.\cite{JenkoPRA2006,StrintziPOP2007} This raises the question whether ion-scale turbulence can also exhibit similar symmetry breaking when there is considerable $k_{x0}$ shift in the linear spectrum. We explore two different examples of drift wave turbulence below: ion temperature gradient instability driven turbulence and trapped electron mode driven turbulence.

\subsection{ITG case}

\begin{figure}
  \includegraphics[width=\columnwidth,clip]{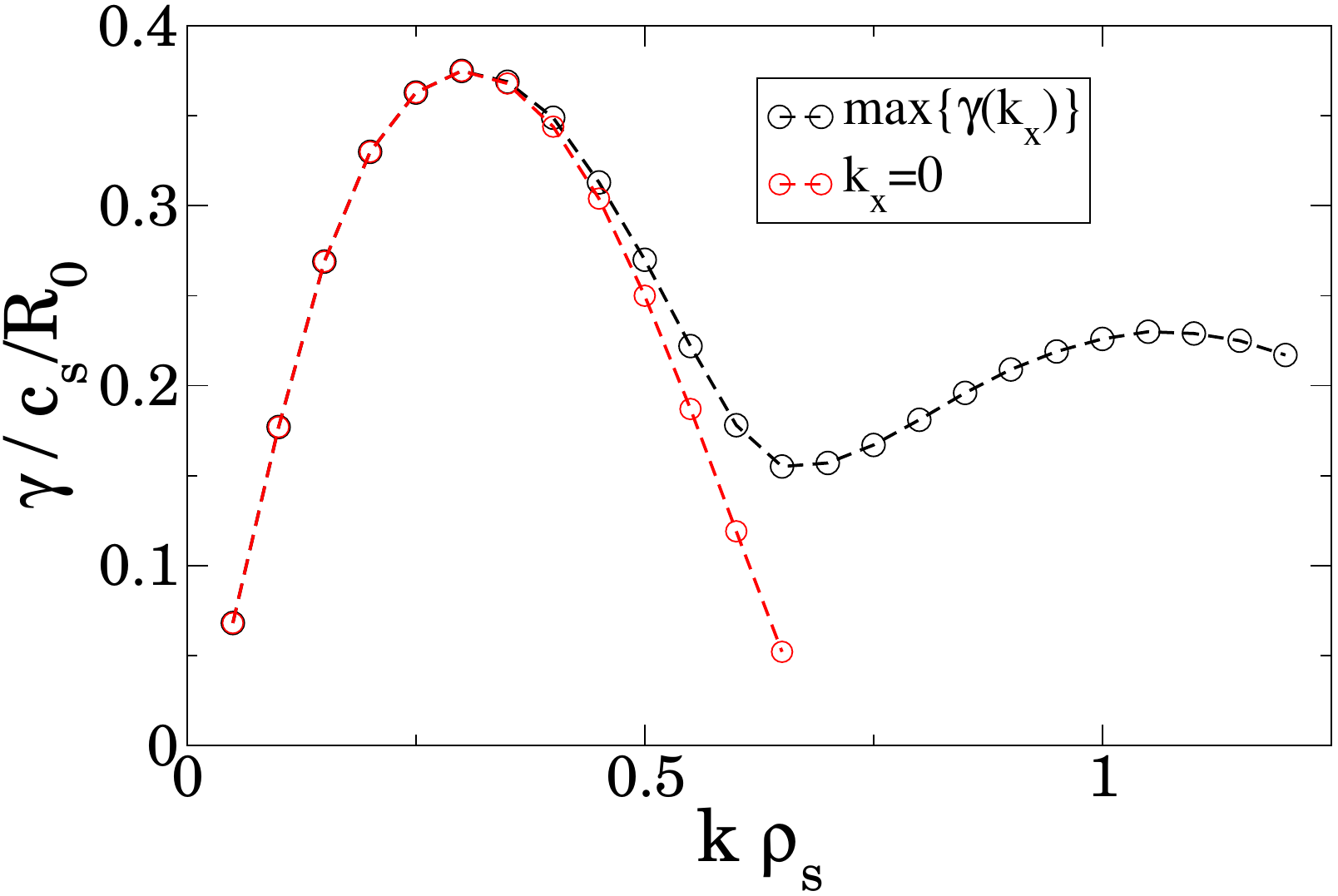}
  \caption{Linear growth rate for the Waltz standard case. At higher $k$ there is a $k_{x0}$ offset, like in our ETG case.}\label{fig:migliano-linear}
\end{figure}

\begin{figure}
  \includegraphics[width=\columnwidth]{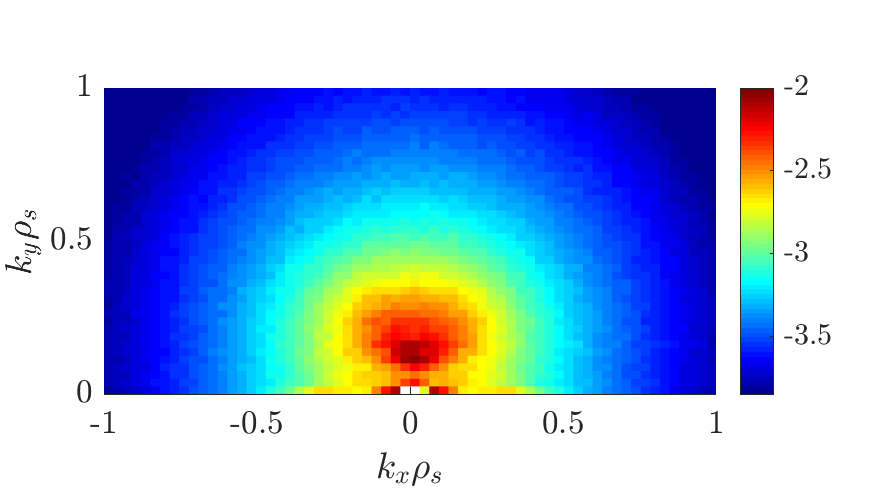}
  \caption{ITG case with $k_{x0}$ offset. Nonlinear spectrum is largely symmetric. The spectrum is more isotropic than in the ETG case due to eddy break-up by zonal modes.}\label{fig:migliano-nonlin}
\end{figure}

The popular ``Waltz standard case'' was shown by Migliano et al.\cite{Migliano2013} to exhibit similar $k_x$ symmetry as the present ETG case, with high-$k$ modes presenting a ballooning angle shift. A similar case was investigated by Singh et al.\cite{SinghPOP2014}. We reproduce the growth rate figure from Migliano et al.\cite{Migliano2013} in figure \ref{fig:migliano-linear} using GENE. The nonlinear spectrum, as show in figure \ref{fig:migliano-nonlin}, does not exhibit broken symmetry in long-term simulations. Shearing rate shows spectral density around similar values of $k_x$ as the $k_y$ amplitude maximum of turbulence, effectively breaking up streamer structures and eliminating asymmetry. This leads to a more isotropic distribution of turbulence compared to the ETG case shown in Fig. \ref{fig:electron_spectrum}.

\subsection{Multi-species TEM case}

\begin{figure}
  \includegraphics[width=\columnwidth,clip]{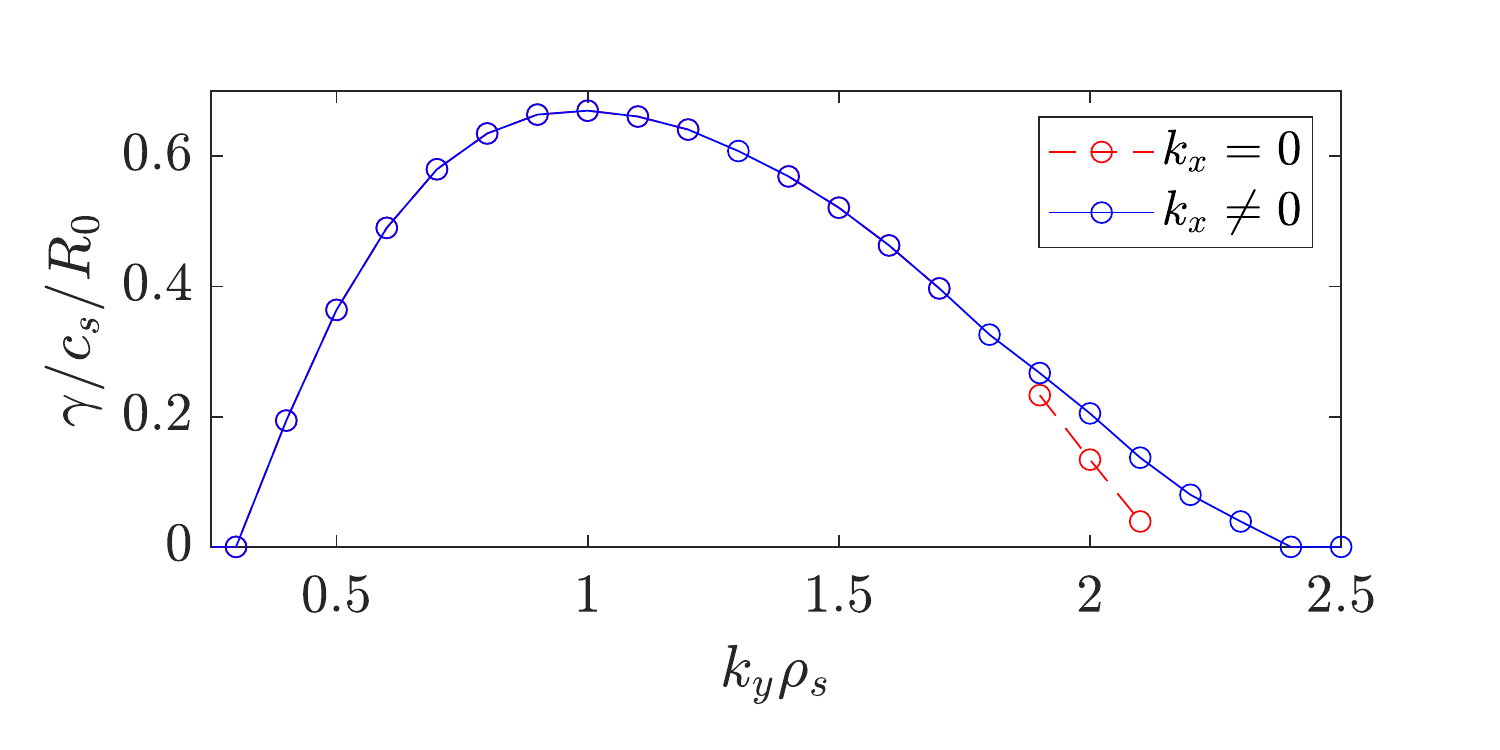}
  \includegraphics[width=\columnwidth,clip]{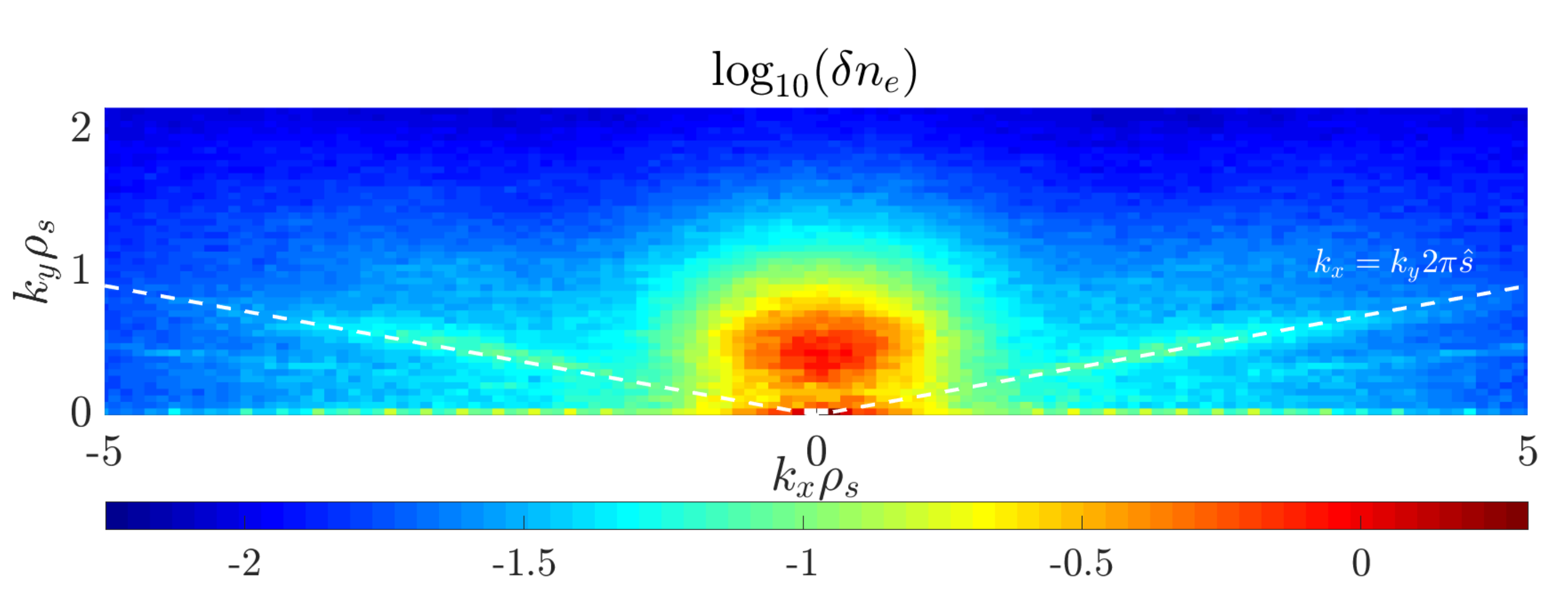}
  \caption{a) Linear growth rate for the 3-species case described in this paper. At low $k$ there is a $k_{x0}$ offset. Nonlinear spectrum (b) is more isotropic than in the ETG case due to eddy break-up by zonal modes. The contribution from rational surfaces is emphasized using the dashed line.}\label{fig:ft2-highd-lin}
\end{figure}

The original case does not have a shift in the ballooning angle for growth rate in trapped electron mode turbulence. We have identified a case with FT-2 parameters for high density discharges where there is a ballooning angle shift, as illustrated by linear growth rate calculations using GENE in Fig. \ref{fig:ft2-highd-lin}. The parameters for this case are modified as follows: $r/a=0.4875$, $q=1.3682$, $\hat{s}=0.8756$, $\omega_n=15.118$, $\omega_{T{e,H}}={19.824,5.4121}$, $T_H/T_e=0.619647$, $T_e=218.3\,$eV, $B_T=2.3\, $T, $n_e=8.8675\cdot10^{19}\,1/$m$^3$, $n_Z/n_i=0.052753$, and the ion species are H and $O^{7+}$ with $T_H/T_z=1$. Eddy break-up is effective at low-$k$, but a very slight asymmetry is seen in the spectrum at high-$k$ (see Fig.~\ref{fig:ft2-highd-lin}c). Shearing occurs at wavelengths comparable to the maximum spectral density of turbulence, as was the case in the ITG case. 

This case has the special feature of a very broad linear spectrum as a function of $k_{x0}$, with sub-dominant high-$k_x$ intermediate $k_y$ modes. This contributes to the anisotropy of the spectrum, observed as a larger width in the $k_x$ spectral density compared to $k_y$. Notable features in the spectrum also include the effect of rational surfaces ($-\,-$), and a mode that is linearly unstable near $(k_x,k_y)=(3.8,0.7)$ with a very broad $k_x$ structure.

\section{Conclusions}

We have identified a case based on FT-2 tokamak experiments that exhibits nonlinear symmetry breaking of the electron temperature gradient instability driven turbulence. The ballooning angle of biased spectrum is affected by the relative directions of magnetic field and plasma current, and is independent of the initial condition that is used to seed the turbulence in simulations or even if a $x$-mirrored restart is performed. Broken symmetry is therefore not spontaneously established, but is related to the radial and binormal wave number components of the drift frequency. Level of transport saturates before the fully broken symmetry is established. Long term simulations of other kinds turbulence (i.e., ITG \& TEM) presented here symmetry breaking is not retained, resulting in a symmetric $(k_x,k_y)$ spectrum for turbulence despite part of the growth-rate spectrum occurring at a finite ballooning angle. At ion scales zonal flows break up streamers while saturating turbulence at scales comparable to it, effectively suppressing the ballooning angle bias. ETG is less sensitive to eddy break-up by shearing, instead saturating through secondary instabilities. The residual streamers can then form spectra of broken symmetry with respect to the radial wave number.

In the literature growth rate spectra are most often reported on only for ballooning angle $\vartheta_0=0$, where in many cases the largest growth rate occurs at ion scales. Some works have taken an interest in modes with $\vartheta_0\neq 0$, but to our knowledge this is the first example of asymmetry in the fully saturated turbulent regime. Even though our TEM case does not exhibit symmetry breaking, it shows significant influence on the spectrum from sub-dominant modes with nonzero ballooning angle that increase the $k_x$ width of the spectral maximum. This may be of importance for experimental diagnostics that are sensitive to radial wave numbers.

The data that supports the findings of this study are available within the article (and its supplementary material).

\noindent{{\bf Acknowledgements}} The work has been supported by the Academy of Finland grants 316088 and 330342. Financial support of the Russian Science Foundation grant 17-12-01110 is acknowledged. Maintenance of the FT-2 tokamak data base is supported by the state contract of the Ioffe institute. CSC -- IT Center for Science is  acknowledged for generous allocation of computational resources for this work. This research was enabled in part by support provided by WestGrid (www.westgrid.ca) and Compute Canada (www.computecanada.ca). The work has also benefited from discussions with Andrei Smolyakov (USASK).


%
%

%


\bibliographystyle{apsrev4-2}
\bibliography{multiscale}

\end{document}